\documentclass{vldb}
\usepackage{verbatim}
\usepackage{graphicx}
\usepackage{balance} 
\usepackage[ruled,linesnumbered]{algorithm2e}
\usepackage{caption}
\usepackage{subcaption}
\usepackage{url}
\usepackage{amsmath}
\usepackage{enumitem}
\usepackage{booktabs}
\usepackage{multirow}
\usepackage{array}

\usepackage{amssymb}
\usepackage{pifont}

\newcommand{\cmark}{\ding{51}}%
\newcommand{\xmark}{\ding{55}}%

\begin{document}

\title{Hippo: A Fast, yet Scalable, Database Indexing Approach}

\numberofauthors{2} 

\author{
\alignauthor
Jia Yu\\
\bigskip
       \affaddr{ Arizona State University}\\
       \affaddr{ 699 S. Mill Avenue, Tempe, AZ}\\
       \email{ jiayu2@asu.edu}\\     
\and       
\alignauthor
       Mohamed Sarwat\\
       \bigskip
      	\affaddr{ Arizona State University}\\
      	\affaddr{ 699 S. Mill Avenue, Tempe, AZ}\\
       \email{ msarwat@asu.edu}\\
}

\maketitle

\begin{abstract}
	
Even though existing database indexes (e.g., B$^+$-Tree) speed up the query execution, they suffer from two main drawbacks: (1)~A database index usually yields 5\% to 15\% additional storage overhead which results in non-ignorable dollar cost in big data scenarios especially when deployed on modern storage devices like Solid State Disk (SSD) or Non-Volatile Memory (NVM). (2)~Maintaining a database index incurs high latency because the DBMS has to find and update those index pages affected by the underlying table changes. This paper proposes Hippo a fast, yet scalable, database indexing approach. Hippo only stores the pointers of disk pages along with light weight histogram-based summaries. The proposed structure significantly shrinks index storage and maintenance overhead without compromising much on query execution performance. Experiments, based on real Hippo implementation inside PostgreSQL 9.5, using the TPC-H benchmark show that Hippo achieves up to two orders of magnitude less storage space and up to three orders of magnitude less maintenance overhead than traditional database indexes, i.e., B$^+$-Tree. Furthermore, the experiments also show that Hippo achieves comparable query execution performance to that of the B$^+$-Tree for various selectivity factors.
\end{abstract}

\section{Introduction}
\label{introduction}
 
A database system (DBMS) often employs an index structure, e.g., B$^+$-Tree~\cite{C79}, to speed up query execution at the cost of additional storage and maintenance overhead. A DBMS user may create an index on one or more attributes of a database table. A created index allows the DBMS to quickly locate tuples without having to scan the whole indexed table. Even though existing database indexes significantly improve the query response time, they suffer from the following drawbacks:

\begin{table}
	\small
	\centering
	\begin{subtable}[t]{1\linewidth}
				\centering
				\vspace{0pt}
				\begin{tabular}{|c |c |c |c|}
					\hline
					TPC-H & Index size & Initialization time  & Insertion time \\
					\hline
					\hline					
					2 GB & 0.25 GB & 30 sec & 10 sec\\
					\hline
					20 GB & 2.51 GB & 500 sec & 1180 sec\\
					\hline
					200 GB & 25 GB & 8000 sec & 42000 sec \\
					\hline
				\end{tabular}
				\caption{B$^+$-Tree overhead}
				\label{table:indexoverhead}
	\end{subtable}\hfill

	\begin{subtable}[t]{1\linewidth}
			\centering
			\vspace{0pt}
			\begin{tabular}{|c | c | c | c|} 
				\hline
				HDD & E-HDD & SSD & E-SSD\\
				\hline
				\hline
			0.04 \$/GB & 0.1 \$/GB & 0.5 \$/GB & 1.4 \$/GB\\
				\hline
			\end{tabular}
			\caption{Storage dollar cost}\label{table:diskprice}
	\end{subtable}

	\caption{Index overhead and storage dollar cost}
	\label{tbl:main}
\end{table}

\begin{itemize}

\item {\bf Indexing Overhead:} Indexing overhead consists of two parts - storage and initialization time overhead. A database index usually yields 5\% to 15\% additional storage overhead. Even though the storage overhead may not seem too high in small databases, it results in non-ignorable dollar cost in big data scenarios. Table~\ref{table:indexoverhead} depicts the storage overhead of a B$^+$-Tree created on the Lineitem table from the TPC-H~\cite{C08} benchmark (database size varies from 2, 20 and 200 GB). Moreover, the storage dollar cost is dramatically amplified when the DBMS is deployed on modern storage devices (e.g., Solid State Drives and Non-Volatile Memory) because they are still more than an order of magnitude expensive than Hard Disk Drives (HDDs) per unit of storage. Table~\ref{table:diskprice} lists the dollar cost per storage unit collected from Amazon.com and NewEgg.com (Enterprise is abbreviated to E). In addition, initializing an index may be a time consuming process especially when the index is created on a large database table. Such high initialization overhead may delay the analysis process (see Table~\ref{table:indexoverhead}). 

\item {\bf Maintenance Overhead:}  A DBMS must update the index after inserting (deleting) tuples into (from) the underlying table. Maintaining a database index incurs high latency because the DBMS has to find and update those index entries affected by the underlying table changes. For instance, maintaining a B$^+$-Tree searches the tree structure and perhaps performs a set of tree nodes splitting or merging operations. That requires plenty of disk I/O operations and hence encumbers the time performance of the entire DBMS in big data scenarios. Table~\ref{table:indexoverhead} shows the B+ Tree insertion overhead (insert 0.1\% records) for the TPC-H Lineitem table.
\end{itemize}

Existing approaches that tackle one or more of the aforementioned drawbacks are classified as follows: {\em (1)~Compressed indexes:} Compressed B$^+$-Tree approaches~\cite{FSV10,GRS98,ZHN+06} reduce index storage overhead but all these methods compromise on query performance due to the additional compression and decompression time. Compressed bitmap indexes also reduce index storage overhead~\cite{GCC+14,LKA10,SW06} but they mainly suit low cardinality attributes which are quite rare. For high cardinality attributes, the storage overhead of compressed bitmap indexes significantly increases~\cite{WOS04}. {\em (2)~Approximate indexes:} Approximate indexing approaches~\cite{AA14,HS05,SYU+00} trade query accuracy for storage to produce smaller, yet fast, index structures. Even though approximate indexes may shrink the storage size, users cannot rely on their un-guaranteed query accuracy in many accuracy-sensitive application scenarios like banking systems or user archive systems. {\em (3)~Sparse indexes:} A sparse index~\cite{BZ98,SE09,SR86,W12} only stores pointers which refer to disk pages and value ranges (min and max values) in each page so that it can save indexing and maintenance overhead. It is generally built on ordered attributes. For a posed query, it finds value ranges which cover or overlap the query predicate and then rapidly inspects the associated few parent table pages one by one for retrieving truly qualified tuples. However, for unordered attributes which are much more common, sparse indexes compromise too much on query performance because they find numerous qualified value ranges and have to inspect a large number of pages. 

This paper proposes Hippo\footnote{\small \url{https://github.com/DataSystemsLab/hippo-postgresql}} a fast, yet scalable, sparse database indexing approach. In contrast to existing tree index structures, Hippo stores disk page ranges (each works as a pointer of one or many pages) instead of tuple pointers in the indexed table to reduce the storage space occupied by the index. Unlike existing approximate indexing methods, Hippo guarantees the query result accuracy by inspecting possible qualified pages and only emitting those tuples that satisfy the query predicate. As opposed to existing sparse indexes, Hippo maintains simplified histograms that represent the data distribution for pages no matter how skew it is, as the summaries for these pages in each page range. Since Hippo relies on histograms already created and maintained by almost every existing DBMS (e.g., PostgreSQL), the system does not exhibit a major additional overhead to create the index. Hippo also adopts a page grouping technique that groups contiguous pages into page ranges based on the similarity of their index key attribute distributions. When a query is issued on the indexed database table, Hippo leverages the page ranges and page summaries to recognize those pages for which the internal tuples are guaranteed not to satisfy the query predicates and inspects the remaining pages. Thus Hippo achieves competitive performance on common range queries without compromising the accuracy. For data insertion and deletion, Hippo dispenses with the numerous disk operations by rapidly locating the affected index entries. Hippo also adaptively decides whether to adopt an eager or lazy index maintenance strategy to mitigate the maintenance overhead while ensuring future queries are answered correctly. 

 

We implemented a prototype of Hippo inside PostgreSQL 9.5. Experiments based on the TPC-H benchmark show that Hippo achieves up to two orders of magnitude less storage space and up to three orders of magnitude less maintenance overhead than traditional database indexes, i.e., B$^+$-Tree. Furthermore, the experiments show that Hippo achieves comparable query execution performance to that of the B$^+$-Tree for various selectivity factors.

The remainder of the paper is structured as follows: In Section~\ref{structure}, we explain the idea of Hippo and show its structure. We demonstrate how to query Hippo swiftly, build Hippo from scratch, and maintain Hippo efficiently in Section~\ref{search},~\ref{build} and~\ref{maintenance}. In Section~\ref{cost}, we provide useful cost estimation for these three scenarios. Extensive experiments and related analysis are included in Section~\ref{experiment}.  We discuss related work then analyze the drawbacks in existing indexes in Section~\ref{relatedwork}. Finally, Section~\ref{conclusion} concludes the paper.

\section{Hippo overview}
\label{structure}

\begin{figure}
	\includegraphics[width=1\linewidth]{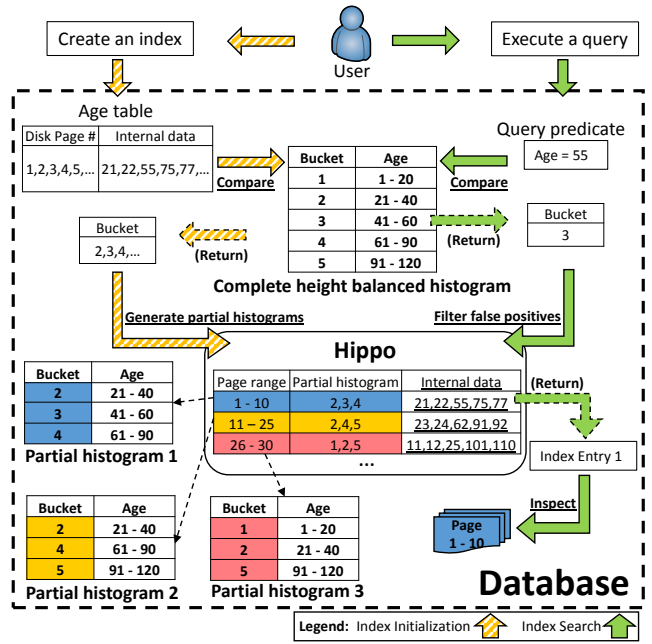}
	\caption{Initialize and search Hippo on age table}
	\label{fig:hbrinsearch}
\end{figure}

This section gives an overview of Hippo. A running example that describes a Hippo index built on an age table is given in Figure~\ref{fig:hbrinsearch}. The figure's right part which depicts how to search Hippo and the left part which shows how to initialize Hippo are explained in Section~\ref{search} and \ref{build} respectively. The main challenges of designing an index are to reduce the indexing overhead in terms of storage and initialization time as well as speed up the index maintenance while still keeping competitive query performance. To achieve that, an index should possess the following two main properties: {\em  (1)~Less Index Entries:} For better storage space utilization, an index should determine and only store the most representative index entries that summarize the key attribute. Keeping too many index entries inevitably results in high storage overhead as well as high initialization time.
{\em (2)~Index Entries Independence:} Index entries of a created index should be independent from each other. In other words, the range of values that each index entry represents should have minimal overlap with other index entries. Interdependence among index entries, like that in a B$^+$-Tree, may lead to overlapped tree nodes traverse during query processing and several cascaded updates during index maintenance.

{\bf Data Structure.} When creating an index, Hippo scans the indexed table and generates histogram-based summaries for disk pages based upon the index key attribute. Afterwards, these summaries are stored by Hippo along with pointers of the pages they summarize. As shown in Figure~\ref{fig:hbrinsearch}, a Hippo index entry consists of the following two components (Internal data of pages is given in the figure only for the ease of understanding):

\begin{itemize}
	
\item {\bf Summarized Page Range:}
The page range (works as a pointer) represents the IDs (i.e., address) of the first and last pages summarized by a certain histogram based summary. DBMS can load particular pages into buffer according to their customized IDs. Hippo is able to summarize more than one contiguous (in terms of physical storage) pages to reduce the overall index size to a great extent (e.g., Page 1 - 10, 11 - 25, 26 - 30). The number of summarized pages (denoted as pages per partial histogram) in each index entry varies. For a certain index attribute, some contiguous pages have very similar content but some are not. Hence, Hippo adopts a page grouping technique that groups contiguous pages into page ranges based on the similarity of their index attribute distributions, using the partial histogram density (explained in Section~\ref{build}). 

\item {\bf Histogram-based Page Summary:} The page summary in each index entry is a partial histogram that represents a subset of the complete height balanced histogram buckets (maintained by the underlying DBMS). Each bucket if exists indicates that at least one of the tuples of this bucket exists in the summarized pages. Each partial histogram represents the distribution of the data in the summarized contiguous pages. Since each bucket of a height balanced histogram roughly contains the same number of tuples, each of them has the same probability to be hit by a random tuple from the table. Hippo leverages this feature to handle data which has various or even skewed distribution. To save storage space, only bucket IDs are kept in partial histograms and partial histograms are stored in a compressed bitmap format. For instance, the partial histogram of the first Hippo index entry in Figure~\ref{fig:hbrinsearch} is stored as $01110$. Each bit, set to 1 or 0, reflects whether the corresponding bucket exists or not.

\end{itemize}

{\bf Main idea.} Hippo solves the aforementioned challenges as follows: (1)~Each index entry summarizes many pages and each only stores two page IDs and a compressed bitmap.(2)~Each page of the parent table is only summarized by one Hippo index entry. Hence, any updates that occur in a certain page only affect a single independent index entry. Finally, during a query, pages whose partial histograms do not have desired buckets are guaranteed not to satisfy certain query predicates and marked as false positives. Thus Hippo only inspects other pages that probably satisfies the query predicate and achieves competitive query performance.

\section{Index search}
\label{search}

The search algorithm runs in three main steps: (1)~Step~1: convert query predicates, (2)~Step~2: filter false positives and (3)~Step~3: inspect possible qualified pages against the query predicate. The search process leverages the index structure to avoid worthless page inspection so that Hippo can achieve competitive query performance.
\begin{algorithm}
	\SetAlgoLined
	\KwData{A given query predicate and Hippo index}
	\KwResult{Qualified tuples}
	Create Bitmap a for the given predicate\;
	\ForEach{bucket of the complete histogram}
	{
			\If{it is hit by the query predicate}
			{
				Set the corresponding bit in Bitmap a to 1\;
			}
	}
	Create Bitmap b for recording all pages\;
	\ForEach{partial histogram}
	{
		\If{it has joint buckets with Bitmap a}
		{
			Set the corresponding bits of the summarized pages in Bitmap b to 1\;
		}	
	}
	\ForEach{page marked as 1 in Bitmap b}{
		{Check each tuple in it against the predicate\;}	
	}
	\caption{Hippo index search}
	\label{algor:search}
\end{algorithm}




\subsection{Convert query predicates}
\label{sec:convquerypred}

\begin{figure}
	\centering
	\includegraphics[width=0.45\textwidth]{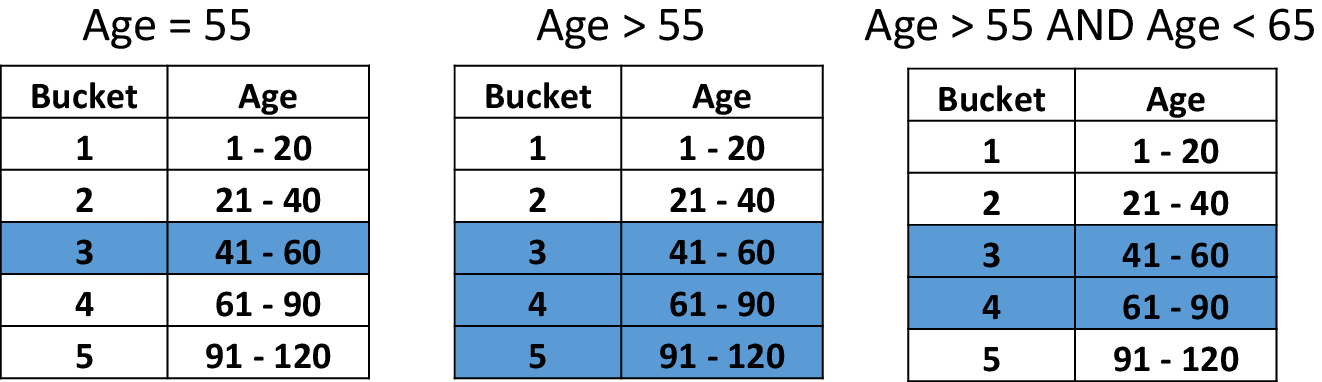}
	\caption{Convert query predicates}
	\label{fig:matchquerypredicate}
\end{figure}


The main idea is to check each partial histogram against the given query predicate for filtering false positives and so speeding up the query. However, as explained in Section~\ref{structure}, partial histograms are stored in bitmap formats without recording value ranges of buckets. Therefore, there has to be an additional step to recover the missing information for each partial histogram on-the-fly or convert the predicate to the bitmap format per query. Obviously, the later one is more efficient.

Any query predicates for a particular attribute can be broken down into atomic units: equality query predicate and range query predicate. Age = 55 is a typical equality query predicate while age \textgreater~55 is a range query predicate. These unit predicates can be combined together by AND operator like age \textgreater~55 AND age \textless 65.


Each unit predicate is compared with the buckets of the complete height balanced histogram (retrieving method is discussed in Section~\ref{build}). A bucket is hit by a predicate if the predicate fully contains, overlaps, or is fully contained by the bucket. Each unit predicate can hit one, at least, or more buckets. For instance, according to the complete histogram in Figure~\ref{fig:hbrinsearch}, bucket 3 whose description is 41 - 60 is hit by age = 55 while bucket 3, 4, and 5 are hit by age \textgreater~55. This strategy is also applicable for the conjunct query predicates. For a conjunct predicate like age \textgreater~55 and age \textless~65, only buckets which are hit by all these unit predicates simultaneously (the joint bucket 3 and 4) are kept as the final result and others are directly discarded. Figure~\ref{fig:matchquerypredicate} shows the hit buckets of three query predicates.
Afterwards, the given query predicate is converted to a bitmap. Each bit in this bitmap reflects whether the bucket has the corresponding ID is hit (1) or not (0). Thus the corresponding bits of all hit buckets are set to 1.

\subsection{Filter false positives}
\label{sec:filtfalsepos}

Filtering false positives is the most important step of Hippo index search. Each Hippo index entry stores a page range and a summary of several contiguous pages but it is very possible that none of these pages in the certain index entry contain the qualified tuples especially for small range queries. This kind of pages and their associated index entries are false positives. This step is to check each partial histogram against the converted query predicate, recognize some false positive pages utmost and finally avoid worthless page inspection on these pages. 

A given query predicate hits one ,at least, or more buckets of the complete histogram. Pages whose partial histograms contain the hit buckets (the corresponding bitmap bits are 1) might have qualified tuples, whereas pages whose partial histograms don't contain these buckets (the corresponding bitmap bits are 0) are guaranteed not to contain qualified tuples. The former kind of pages are possible qualified pages. In contrast, the later kind of pages are false positives and excluded from the next step - inspect possible qualified pages. The straight way to find false positive pages is to do a nested loop between each partial histogram and the converted query predicate to find the joint buckets.

\begin{figure}[t]
	\centering
	\includegraphics[width=0.5\linewidth]{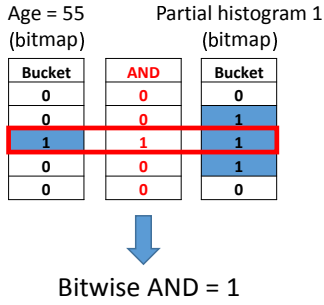}
	\caption{Bitwise AND two bitmaps to find joint buckets}
	\label{fig:filterhistograms}
\end{figure}

Interestingly, because both of partial histograms and the converted query predicate are in bitmap format, the nested loop can be accelerated by bitwise 'AND'ing the bytes from both sides, aka bit-level parallelism. If bitwise 'AND'ing the two bytes from both sides returns 1, that means there are joint buckets between the query predicate and the partial histogram. Thus the pages are possible qualified pages. Figure~\ref{fig:filterhistograms} provides an example of how to perform a bitwise AND using the same data in Figure~\ref{fig:hbrinsearch}.

\subsection{Inspect possible qualified pages}
\label{sec:inspectpages}

The previous step recognizes many false positive pages and excludes them from possible qualified pages. However, one fact is that not all false positives can be detected by the previous step. Possible qualified pages still may contain false positives and this is why they are called "possible". This step is to inspect the tuples in each possible qualified pages and retrieve the qualified tuples directly.

IDs of possible qualified pages are recorded in a separate bitmap. Each bit in this bitmap is mapped to the page at the same position in the parent table. For instance, the bit at position 1 in the bitmap is mapped to the page ID 1 of the parent table. The value (1 or 0) of this bit reflects whether the associate page is a possible qualified page or not.

Hippo has to inspect all of the possible qualified pages recorded in the bitmap against the query predicate one by one because every retained page from the previous step is possible to contain qualified tuples. The only way to inspect these possible qualified pages is to traverse them and check each tuple in each page one by one. Qualified tuples are returned to the DBMSs.

Algorithm~\ref{algor:search} shows the three steps of Hippo index search. The right part of Figure~\ref{fig:hbrinsearch} describes how to search Hippo index using a certain query predicate. Firstly, Hippo finds query predicate age = 55 hits bucket 3. And the first one of the three partial histograms nicely contains bucket 3. Thus only the disk pages 1 - 10 are selected as possible qualified pages which need further inspection. It is also worth noting that these partial histograms summarize different number of pages.

\section{Index initialization}
\label{build}
Hippo performs three main steps to initialize itself: (1)~Retrieve a complete histogram, (2)~Generate partial histograms, and (3)~Group similar pages into page ranges, described as follows. 

\begin{algorithm}
	\SetAlgoLined
	\KwData{Pages of a parent table}
	\KwResult{Hippo index}
	Create a working partial histogram (in bitmap format)\;
	Set StartPage = 1 and EndPage = 1\;
	\ForEach{page}{
		Find distinct buckets hit by its tuples\;
		Set associated bits to 1 in the partial histogram\;
		\uIf{the working partial histogram density \textgreater~threshold}
		{
		Store the partial histogram and the page range (StartPage and EndPage) as an index entry\;
			Create a new working partial histogram\;
			StartPage = EndPage + 1\;
			EndPage = StartPage\;
		}
		\Else
		{EndPage = EndPage + 1\;}	
	}
	\caption{Hippo index initialization}
	\label{algor:summarizemorepages}
\end{algorithm}

\subsection{Retrieve a complete histogram}
\label{retrievehistogram}

Histograms used in Hippo include a complete height balanced histogram and many partial histograms. A complete height balanced histogram represents the distribution of all tuples and already exists in DBMSs. Respectively, a partial histogram, as a subsection, only contains partial buckets from the complete histogram. Therefore, for generating any partial histograms, a complete histogram should be retrieved at the first priority. Full-fledged functions for retrieving a complete histogram exist in any DBMSs. Detailed explanation for these functions is omitted in this paper since it is not our focus. We also assume that the complete histogram is not changed at any time because the global distribution of the parent table will not be affected even if some local updates are performed.
 
The resolution of the complete histogram (denoted as histogram resolution) is adjustable. A complete histogram is considered as higher resolution if it contains more buckets. The resolution of partial histograms is consistent with their complete histograms technically. It is apparent that a complete histogram will have larger physical size if it has higher resolution and, accordingly, the numerous partial histograms are also physically larger than the low resolution ones. On the other hand, the histogram resolution also affects Hippo query time. The cost estimation section will further discuss this issue.

\subsection{Generate partial histograms}
\label{generatepatialhistogram}

A partial histogram only contains some buckets from the complete histogram. It is used to represent the distribution of parent tuples in one or many disk pages. In other words, people can get an approximate overview from the partial histogram of these pages: What values might lie in these pages and what do not. These partial histograms are able to help Hippo to recognize false positives utmost and avoid worthless page inspection. We explain how to generate a partial histogram for each disk page in this section.

Generating partial histograms traverses all disk pages of the parent table from the start to end. For each page, a nested loop passes through each tuple in this page. The specified attribute value is extracted from each tuple and compared with the complete histogram (using a binary search). Buckets hit by tuples are kept for this page and then compose a partial histogram. A partial histogram only contains distinct buckets. For instance, there is a group of age values like the first entry of Hippo shown in Figure~\ref{fig:hbrinsearch}: 21, 22, 55, 75, 77. Bucket 2 is hit by 21 and 22, bucket 3 is hit by 55 and bucket 4 is hit by 77. Therefore, the partial histogram for these values is just as partial histogram 1 in Figure~\ref{fig:hbrinsearch}.


Shrinking the physical size of partial histograms is desirable. The basic idea is to drop all bucket value ranges and only keep bucket IDs. Hippo in Figure~\ref{fig:hbrinsearch} shows the effect. Actually, as mentioned in Section~\ref{structure}, dropping value range information does not impact much on the index search. To further shrink the size, storing bucket IDs in integer type (4 bytes or more) is also considered as an overhead. Bitmap format storage is a better choice to bypass this overhead. Each partial histogram is stored as a bitmap. Each bit in a bitmap stands for a bucket at the same position in a complete histogram. Bit value 1 means the associated bucket is hit and kept in this partial histogram while 0 means the associated bucket is not included. Bitmap compression is introduced to Hippo as well. The partial histogram in a bitmap format can be compressed by any existing compression techniques. The time of compressing and decompressing partial histograms is ignorable in contrast to that of inspecting possible qualified pages.

\subsection{Group similar pages into page ranges}
\label{summarizemorepages}

Generating a partial histogram per one disk page is as easy as that in Section~\ref{generatepatialhistogram}. However, for some contiguous pages which have similar data, it is a waste of storage. Grouping them together as many as possible and merging their partial histograms into one larger partial histogram (in other words, summarizing more pages within one partial histogram) can make Hippo more efficient. On the other hand, users may want to shrink Hippo physically to a greater extent. For example, if a partial histogram can summarize 10 pages in one go, the new Hippo size will be much smaller. Grouping more pages into one page range and summarizing them with just one partial histogram are expected and practical as well. 

Yet, this is not saying that all pages should be grouped together and summarized by one merged partial histogram. As more and more pages are summarized, this partial histogram contains more and more buckets until all buckets from the complete histogram are included. At this moment, this partial histogram becomes a complete histogram and covers any possible query predicates. That means this kind of partial histograms is unable to help Hippo to filter the false positives and the disk pages summarized by this partial histogram will be always treated as possible qualified pages. 

One strategy is to group a fixed number of contiguous pages per range/partial histogram. Yet, this strategy is not suitable if some contiguous pages in a certain area have much more similar data distribution than other areas. Lacking the awareness of data distribution cannot reduce storage overhead smartly. Under this circumstance, it is better to let Hippo group more pages together in this area and group less pages together in other areas dynamically. For instance, assume original pages per partial histogram is 100. If there are 1000 out of 10000 disk pages and the tuples in these 1000 pages are exactly same, a better way to shrink the index size is to set the P from 100 to 1000 for grouping/summarizing these 1000 pages into one range/partial histogram and change it back to 100 for other 9000 pages. 

A terminology - partial histogram density is introduced here. The density of a partial histogram is the percentage of kept buckets in the total buckets of a complete histogram. The complete histogram has a density value of 1. The definition can be formalized as follows:
\begin{align*}
Partial~histogram~density~=~\frac{Buckets_{partial~histogram}}{Buckets_{complete~histogram}}
\end{align*}
This density has an important phenomenon that, for a group of contiguous pages, their merged partial histogram density will be very low if these pages are very similar, vice versa. Therefore, a partial histogram with a certain density may summarize more pages if these contiguous pages have similar data,  vice versa. Making use of this phenomenon enables Hippo to dynamically group pages and merge partial histograms into one. In addition, it is understandable that a lower density partial histogram (summarizes less pages) has the high probability to be recognized as false positives so that speed up queries.

User can easily set a same density for all partial histograms as a threshold. Each partial histogram can automatically decide how many pages it should summarize. Algorithm~\ref{algor:summarizemorepages} depicts how to initialize a Hippo and summarize more pages within one partial histogram with the help of the partial histogram density. The basic idea is that new pages will not be summarized into a partial histogram if its density is larger than the threshold and a new partial histogram will be created for the following pages. 

Figure~\ref{fig:hbrinsearch}'s left part depicts how to initialize a Hippo on the age table with a partial histogram density 0.6. All of the tuples are compared with the complete histogram and IDs of distinct buckets hit by tuples are generated as partial histograms along with page range. 

So far, as Figure~\ref{fig:hbrinsearch} shows, each entry in a Hippo index has the following content: a partial histogram in compressed bitmap format and two integers stand for the first and last pages summarized by this histogram (summarized page range). Each entry is serialized and stored on disk.

\section{Index maintenance}
\label{maintenance}

Inserting (deleting) tuples into (from) the indexed table requires maintaining the index to ensure that the DBMS can retrieve the correct set of tuples that match the query predicate. However, the overhead of maintaining the index quite frequently may preclude system scalability. This section explains how Hippo handles updates. 

 \begin{algorithm}[t]
 	\SetAlgoLined
 	\KwData{A new inserted tuple belongs to Page a}
 	\KwResult{Updated Hippo}
	Find the bucket hit by the inserted tuple;
	
 	Locate a Hippo index entry which summarizes Page a;
 	
 	\uIf{one index entry is located}
 	{
 		Retrieve the associated Hippo index entry\;
 		Update the retrieved entry if necessary;

 	}
	\Else{
 		Retrieve the Hippo entry summarizes the last page;
 		
  		\uIf{the partial histogram density \textless~threshold}
  		{Summarize Page a into the retrieved entry;
  		}
  		\Else{Summarize Page a into a new entry;}	 			
 	}		
 	\caption{Update Hippo for data insertion}
 	\label{algor:hbrininsert}
 \end{algorithm}

\subsection{Data insertion}

Hippo should instantly update or check the index at least after inserting one record into the indexed table. Otherwise, all subsequent queries might miss the newly inserted tuple since it is not reflected by the index. Therefore, Hippo adopts an eager update strategy when a new tuple is inserted.
Data insertion may change the physical structure of a table. The new tuple may belong to any pages of the indexed table.
The insertion procedure (See Algorithm~\ref{algor:hbrininsert}) performs the following steps: (1)~Find buckets hit by the new tuple, (2)~Locate the affected index entry, and (3)~Update the index entry if necessary.

{\bf Step~1:~Find buckets hit by the new tuple:} 
Similar with some steps of generating partial histogram in index initialization, after retrieving the complete histogram, the newly inserted tuple is checked against it using a binary search and a bucket hit by this new tuple is found.

{\bf Step~2:~Locate the affected index entry:}
The new tuple has to belong to one page in this table. This page may be a new one which has not been summarized by any partial histograms before or an old one which has been summarized. However, because the numbers of pages summarized by each histogram are different, searching Hippo index entry to find the one contains this target page is inevitable. From the perspective of disk storage, in a Hippo, all partial histograms are stored on disk in a serialized format. It will be extremely time-consuming if every entry is retrieved from disk, de-serialized and checked against the target page. Therefore, a binary search on Hippo index entries is a good choice. (This search actually leverages the index entries sorted list explained in Section~\ref{sortedlist}.) 

{\bf Step~3:~Update the index entry:}
If the new tuple belongs to a new page not summarized by any Hippo index entries and the density of Hippo partial histogram which summarizes the last disk page is smaller than the density threshold set by users, this new page will be summarized into this partial histogram in the last index entry otherwise a new partial histogram will be created to summarize this page and stored in a new Hippo index entry. For a new tuple belongs to pages already summarized by Hippo, the partial histogram in the associated index entry will be updated if the inserted tuple hits a new bucket. 

It is worth noting that: (1)~Since the compressed bitmaps of partial histograms may have different size, the updated index entry may not fit the space left at the old location. Thus the updated one may be put at the end of Hippo. (2)~After some changes (replacing old or creating new index entry) in Hippo, the corresponding position of the sorted list needs to be updated.

\subsection{Data deletion}

The eager update strategy is not highly desired for data deletion. Hippo still ensures the correctness of queries even if it doesn't update itself at all after deleting tuples from a table. This benefits by inspecting possible qualified pages in index search. Pages used to have qualified tuples might be still marked as possible qualified pages but they are discarded after being inspected against the query predicates. A periodic update or bulk update will be a good choice here. 

For data deletion, Hippo adopts a lazy update strategy that maintains the index after a bulk of delete operations. In such case, Hippo traverses each index entry from the start to end. For each index entry, Hippo inspects the header of each summarized page for seeking notes made by DBMSs (e.g., PostgreSQL makes notes in page headers if data is removed from pages). Hippo re-summarizes the entire index entry instantly within the original page range if data deletion on one page is detected. The re-summarization follows the same steps in Section~\ref{build}. It is worth noting that this updated Hippo index entry is not leading to the update on the sorted list because the updated partial histogram, having same or less buckets, can obtain same or less compress bitmap size and the new index entry certainly fits the old space.

\subsection{Index Entries Sorted List}\label{sortedlist}

\begin{figure}
	\centering
	\includegraphics[width=0.45\textwidth]{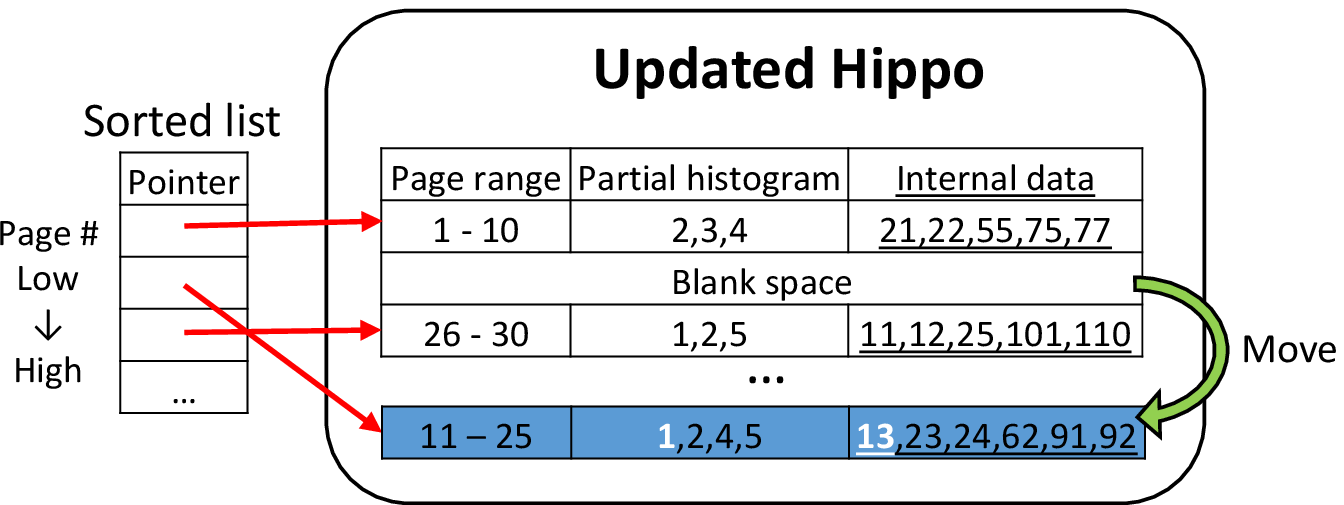}
	\caption{Hippo Index Entries Sorted List}
	\label{fig:pointerarray}
\end{figure}

When a new tuple is inserted, Hippo executes a fast binary search (according to the page IDs) to locate the affected index entry and then updates it. Since the index entries are not guaranteed to be sorted based on the page IDs (noted in data insertion section), an auxiliary structure for recording the sorted order is introduced to Hippo. 

The sorted list is initialized after all steps in Section~\ref{build} with the original order of index entries and put at the first several index pages of Hippo. During the entire Hippo life time, the sorted list  maintains a list of pointers of Hippo index entries in the ascending order of page IDs. Actually each pointer represents the fixed size physical address of an index entry and these addresses can be used to retrieve index entries directly. That way, the premise of a binary search has been satisfied. Figure~\ref{fig:pointerarray} depicts the Hippo index entries sorted list. Index entry $2$ in Figure~\ref{fig:hbrinsearch} has a new bucket ID $1$ due to a newly inserted tuple in its internal data and hence this entry becomes the last index entry in Figure~\ref{fig:pointerarray}. The sorted list is still able to record the ascending order and help Hippo to perform a binary search on the index entries. In addition, such sorted list leads to slight additional maintenance overhead: Some index updates need to modify the affected pointers in the sorted list to reflect the new physical addresses.

\section{Cost Estimation}
\label{cost}

This section gives a detailed cost estimation of Hippo. 
We first provide an accurate query time cost model which assists the DBMS query optimizer in picking an efficient query execution plan. 
Estimating the storage overhead of an index can also facilitate better disk space management and planning.
Index initialization certainly consumes a large chunk of time. Similarly, index maintenance can present a significant time overhead in any write-intensive application. Both of them should be carefully estimated.
 
Table~\ref{table:terms} summarizes the main notations we use to derive the cost model.
Given a database table $R$ with $Card$ number of tuples (i.e., cardinality) and average number of tuples per disk page equal to $pageCard$, a user may create a Hippo index on attribute $a_i$ of $R$. When initializing the index, Hippo sets the complete histogram resolution to $H$ (it has $H$ buckets in total) and the partial histogram density to $D$. Assume that each Hippo index entry summarizes $P$ indexed table pages (in terms of pages)/ $T$ tuples (in terms of tuples). $P$ and $T$ vary for each index entry. Queries executed against the index have average selectivity factor $SF$. 
 


\bgroup
\def\arraystretch{1}
\begin{table}
	\small
	\centering
	\begin{tabular}{|c | l|}
		\hline
		\multicolumn{1}{|c |}{Term} & 
		\multicolumn{1}{c|}{Definition}\\
		\hline
		\hline
		H & \parbox{6.4cm}{Complete histogram resolution which means the number of buckets in this complete histogram}\\[2ex]
		\hline
		D & \parbox{6.4cm}{Partial histogram density which is an user supplied threshold}\\[2ex]
		\hline
		P & \parbox{6.4cm}{Pages summarized by one partial histogram for a certain attribute}\\[2ex]
		\hline
		T & \parbox{6.4cm}{Tuples summarized by one partial histogram for a certain attribute}\\[2ex]
		\hline
		Card & \parbox{6.4cm}{Cardinality (the total number of tuples) of the indexed table}\\[2ex]
		\hline
		pageCard & \parbox{6.4cm}{Number of tuples in each page of the indexed table}\\[1ex]
		\hline
		SF & \parbox{6.4cm}{Query selectivity factor = $\frac{Query~output}{Query~input}$ * 100\% }\\[1ex]
		
		\hline
	\end{tabular}
	\caption{Notations used in Cost Estimation}
	\label{table:terms}
\end{table}
\egroup

\subsection{Query time}


The first step of Hippo index search is to traverse Hippo index entries. Pages in each index entry are likely to be selected for further inspection if their associated partial histogram has joint buckets with the query predicate. Determining the probability of having joint buckets contributes to the query time cost estimation.

\begin{figure}[t]
	\centering
	\includegraphics[width=0.4\textwidth]{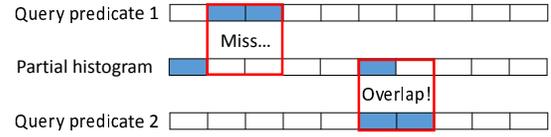}
	\caption{Visualize how to filter false positives}
	\label{fig:cost}
\end{figure}

For the ease of presentation, Figure~\ref{fig:cost} visualizes the procedure of filtering false positives according to their partial histograms. Partial histogram density ($D$) of this index is 0.2. The complete histogram constitutes of $10$ buckets in total ($H=10$). Assume the indexed table's tuples follow an uniform distribution based upon the key attribute. Let the query selectivity factor ($SF$) be 20\%. In Figure~\ref{fig:cost}, buckets hit by the query predicates and the partial histogram are represented in a bitmap format. According to this figure, the partial histogram misses a query predicate if the highlighted area of the predicate falls into the blank area of the partial histogram, whereas a partial histogram is selected if the predicate does not fall completely into the blank area of the histogram. In other words, the probability of a partial histogram having joint buckets with a predicate depends on how likely a predicate doesn't fall into the blank area of a partial histogram. The probability is determined by the formula given below (The terms are defined in Table~\ref{table:terms}):

\begin{align}
Prob &= (Buckets~hit~by~a~query~predicate) * D\notag     \\
&= (SF * H) * D
\end{align}

To be precise, $Prob$ follows a piecewise function as follows:

\[ Prob = \begin{cases} 
(SF * H) * D & S * H\leqslant \frac{1}{D} \\
1 & SF * H > \frac{1}{D}
\end{cases}
\]

$SF * H$ $\in$ \{1, 2, 3, 4, ...\}

$SF * H$ should be no smaller than 1 no matter how small $SF$ is. Because the query predicate at least hits one bucket of the complete histogram. Therefore, the probability in Figure~\ref{fig:cost} is $20\% \times 10 \times 0.2 = 40\%$. That means pages summarized by each index entry have $40$\% probability to be selected as possible qualified pages. 
Given the aforementioned discussion, we observe the following:

{\bf Observation 1:}~When $SF$ and $H$ are fixed, the smaller $D$ is, the smaller $Prob$ is.

{\bf Observation 2:}~When $H$ and $D$ are fixed, the smaller $SF$ is, the smaller $Prob$ is.

{\bf Observation 3:}~When $SF$ and $D$ are fixed, the smaller $H$ is, the smaller $Prob$ is. 

In fact, the probability given above is equal to the percentage of inspected tuples in all tuples. In addition, considering that Hippo index entries are much less than the inspected tuples of the parent table, the total query time cost estimation is mainly decided by the time spent on inspecting possible qualified pages. 
Thus, the query time cost estimation (in terms of disk I/O) can be concluded as follows:

\begin{align}
Query~time &= (Prob * Card)
\end{align}

If we substitute $Prob$ with its piecewise function, the query time cost is as follows:

\[ Query~time = \begin{cases} 
(SF * H) * D * Card & SF * H \leqslant \frac{1}{D} \\
Card & SF * H > \frac{1}{D}
\end{cases}
\]

$SF * H$ $\in$ \{1, 2, 3, 4, ...\}

\subsection{Indexing overhead}
\label{creationestimation}

Indexing overhead which consists of storage overhead and initialization time highly hinges on the number of index entries in an index. The more index entries there are, the more disk writes and storage space an index costs. B$^+$-Tree and other indexes take huge disk space and time for storing their substantial nodes one by one.

The first problem in estimating the number of Hippo index entries is that: how many disk pages ($P$) are summarized by one partial histogram in general? Or, how many tuples ($T$) are checked against the complete histogram for generating one partial histogram? Interestingly, this problem is very similar with Coupon Collector's Problem\cite{FGT92}. This problem can be described like that: "A vending machine sells $H$ types of coupons (a complete histogram with $H$ buckets). Alice is purchasing coupons from this machine. Each time (each tuple) she can get a random type coupon (a bucket) but she might already have a same one. Alice keeps purchasing until she gets $D * H$ types of coupons (distinct buckets). How many times ($T$) does she need to purchase?"


Therefore, the expectations of $T$ and $P$ are determined by the following formulas (The terms are  defined in Table~\ref{table:terms}):

\begin{align}
T&= \frac{H}{H} + \frac{H}{H-1} + \frac{H}{H - 2} + ... + \frac{H}{H - D * H +1}\notag\\
&=H * (\frac{1}{H} + \frac{1}{H-1} + ... +\frac{1}{H - D*H+1})\label{estrows}\\
P &=\frac{T}{pageCard}
\end{align}

$D$ $\in$ [$\frac{pageCard}{H}$, 1]

The product of $D * H$ is the actual number of buckets in each partial histogram. This value should be no smaller than the tuples per disk page ($pageCard$) in case that each tuple in a certain page hit one unique bucket.

For instance, in a Hippo, the complete histogram has 1000 buckets in total and the partial histogram density is 0.1. The expectation of $T$ and $P$ will be 105.3 and $\frac{105.3}{pageCard}$ respectively. That means each partial histogram may summarize $\frac{105.3}{pageCard}$ pages under this circumstance. In another example, if the total number of buckets is 10000 and the density is 0.2, $T$ and $P$ will be 2230 and $\frac{2230}{pageCard}$ correspondingly.

After being aware of the expectation of the number of $P$, it is not hard to deduce the approximate number of index entries in a Hippo. Thus the estimation of Hippo index entries number is Formula~\ref{rows}. If we substitute $T$ with their mathematical expectations in Formula~\ref{estrows} and Formula~\ref{rows} will be changed to Formula~\ref{hipporows}. Hippo index size is equal to the product of the number of index entries and the size of one entry which roughly depends on each partial histogram size (in compressed bitmap format).

\begin{align}
Hippo~index~entries~&= \frac{Card}{T}\label{rows}\\
&=\frac{Card}{H * (\frac{1}{H} + \frac{1}{H-1} + ... + \frac{1}{H - D*H+1})}\label{hipporows}
\end{align}

D $\in$ [$\frac{pageCard}{H}$, 1]

Some observations can be obtained from Formula~\ref{hipporows}:

{\bf Observation 1}~For a certain $H$, the higher $D$ there is, the less Hippo index entries there are.

{\bf Observation 2}~For a certain $D$, the higher $H$ there is, the less Hippo index entries there are. Meanwhile, the size of each Hippo index entry is increasing with the growth of the complete histogram resolution.

Index initialization time hinges on the number of disk I/Os because it takes much more time than memory I/Os. In general, the initialization time is composed of two parts: retrieve parent tuples one by one and write index entry to disk one by one. Accordingly, Hippo initialization time can be deduced as follows:

\begin{align}
Hippo~initialization~time &= Card + Hippo~index~entries
\end{align}
The number of Hippo index entries mentioned in the formula above can be substituted by its mathematical expectation in Formula~~\ref{hipporows}.

\subsection{Maintenance time}

{\bf Data insertion.} Hippo updates itself eagerly for data insertion so that this operation is relatively time sensitive. There are five steps cost disk I/Os in this update: retrieve the complete histogram, locate associated Hippo index entry, retrieve the associated index entry, update the index entry (if necessary) and update the mapped sorted list element. It is not hard to conclude that locating the associated index entry completes in log(Hippo index entries) I/O times, whereas other four steps are able to accomplish their assignments in constant I/O times. Thus the data insertion time cost estimation model is summarized as follows under different conditions:

\begin{align}
Data~insert~time &= log(Hippo~index~entries) + 4\label{insertioncost}
\end{align}
Hippo index entries mentioned in Formula~\ref{insertioncost} can be substituted by its mathematical estimation in Formula~\ref{hipporows}.

{\bf Data deletion.} Hippo updates itself lazily for data deletion so that it is hard to finalize a general estimation model. However, it is recommended that do not update Hippo for data deletion too frequently because Hippo will re-traverse and re-summarize all disk pages summarized by one Hippo index entry once it detects that one disk page has data deletion. This algorithm is more suitable for bulk deletion and lazy update strategy. 


\begin{figure*}[t]
	\centering
	\begin{subfigure}{0.32\textwidth}
		\centering
		\includegraphics[width=0.95\linewidth]{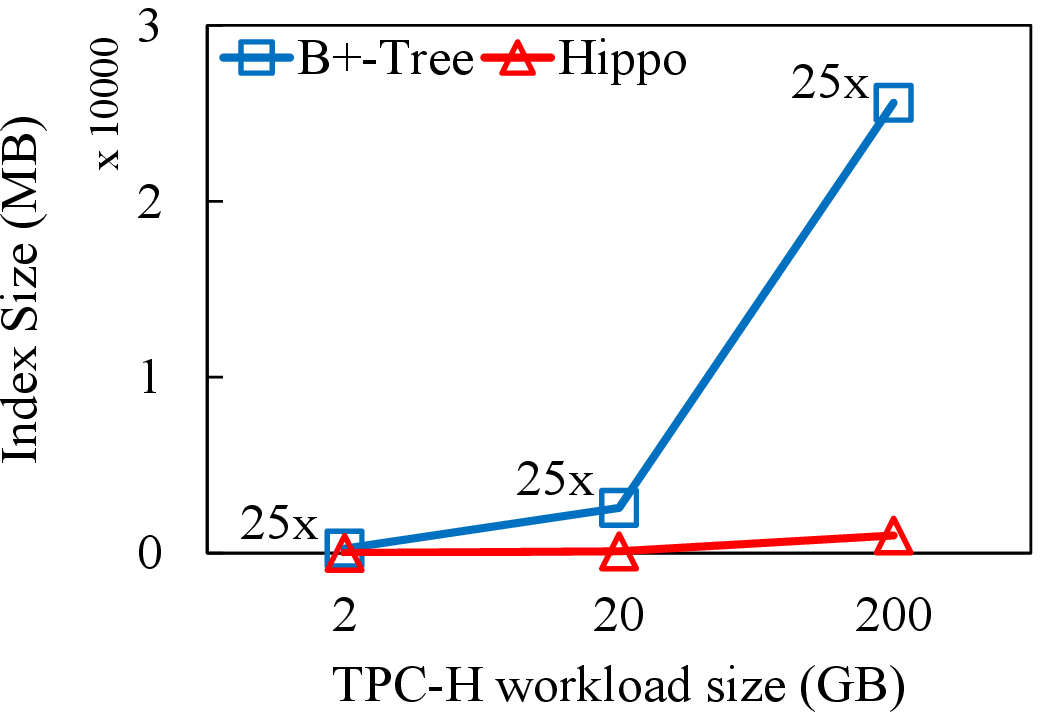}
		\caption{Index size}
		\label{fig:indexsize}
	\end{subfigure}
	~
	\begin{subfigure}{0.32\textwidth}
		\centering
		\includegraphics[width=0.95\linewidth]{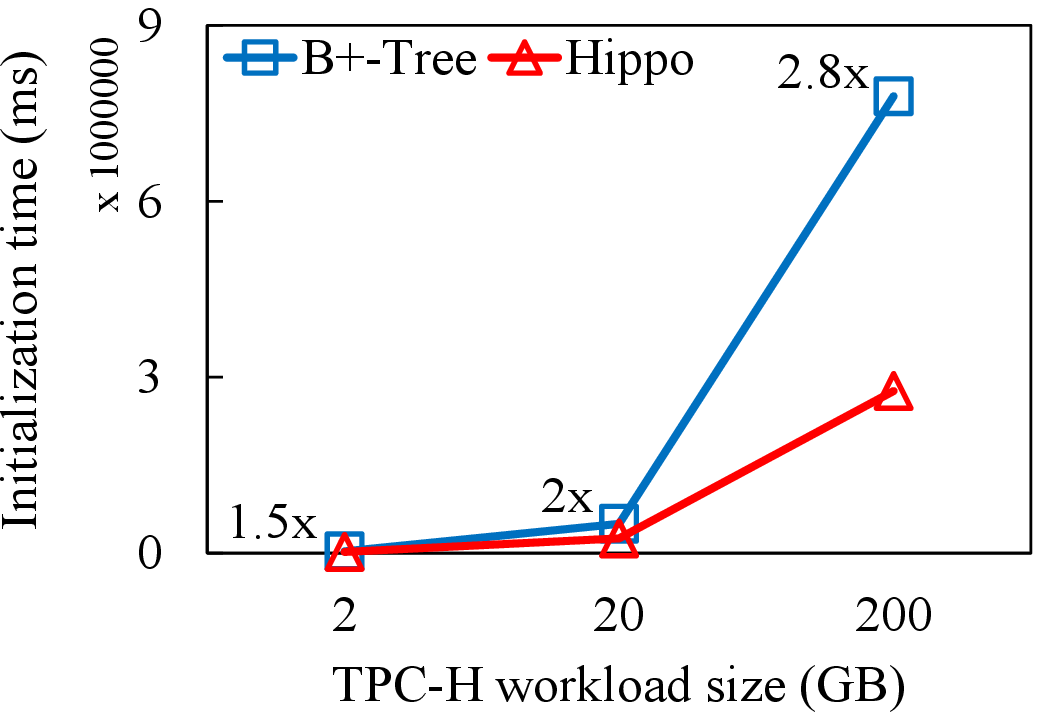}
		\caption{Index initialization time}
		\label{fig:indexconstructiontime}
	\end{subfigure}
	~
	\begin{subfigure}{0.32\textwidth}
		\centering
		\includegraphics[width=0.95\linewidth]{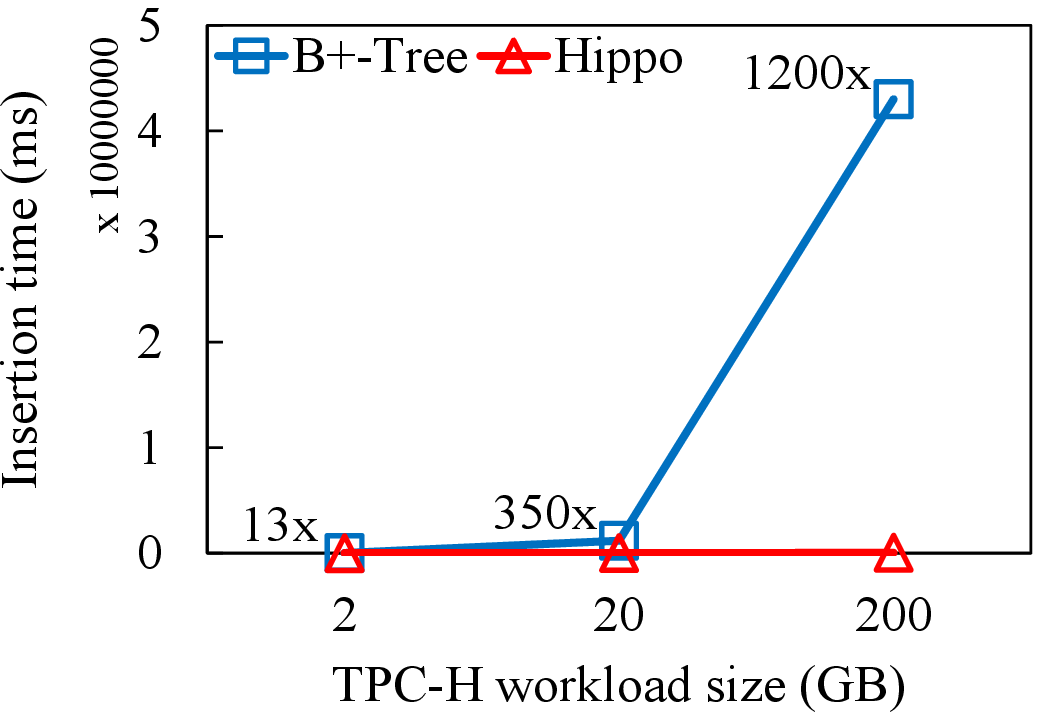}
		\caption{Index insertion time}
		\label{fig:insertamount}
	\end{subfigure}
	\caption{Index overhead on different TPC-H workload size}\label{fig:indexsizecreation}
	
\end{figure*}
\begin{figure*}[t]
	
	\centering
	\begin{subfigure}{0.32\textwidth}
		\centering
		\includegraphics[width=0.95\linewidth]{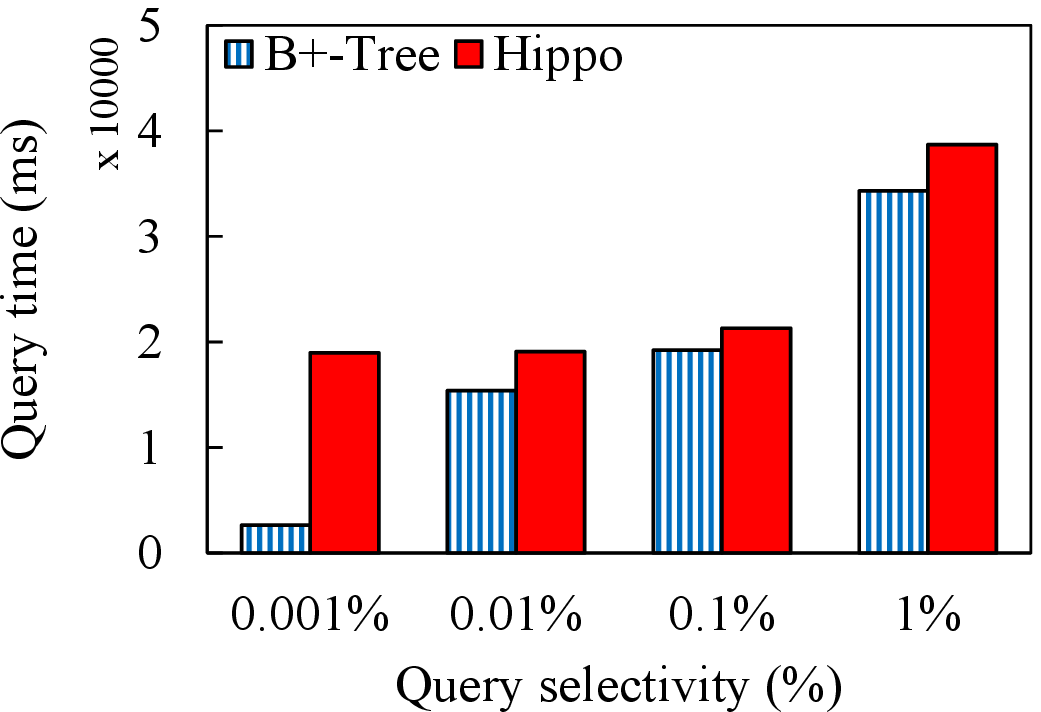}
		\caption{2 GB}
		\label{fig:smallselectivity}
	\end{subfigure}
	~
	\begin{subfigure}{0.32\textwidth}
		\centering
		\includegraphics[width=0.95\linewidth]{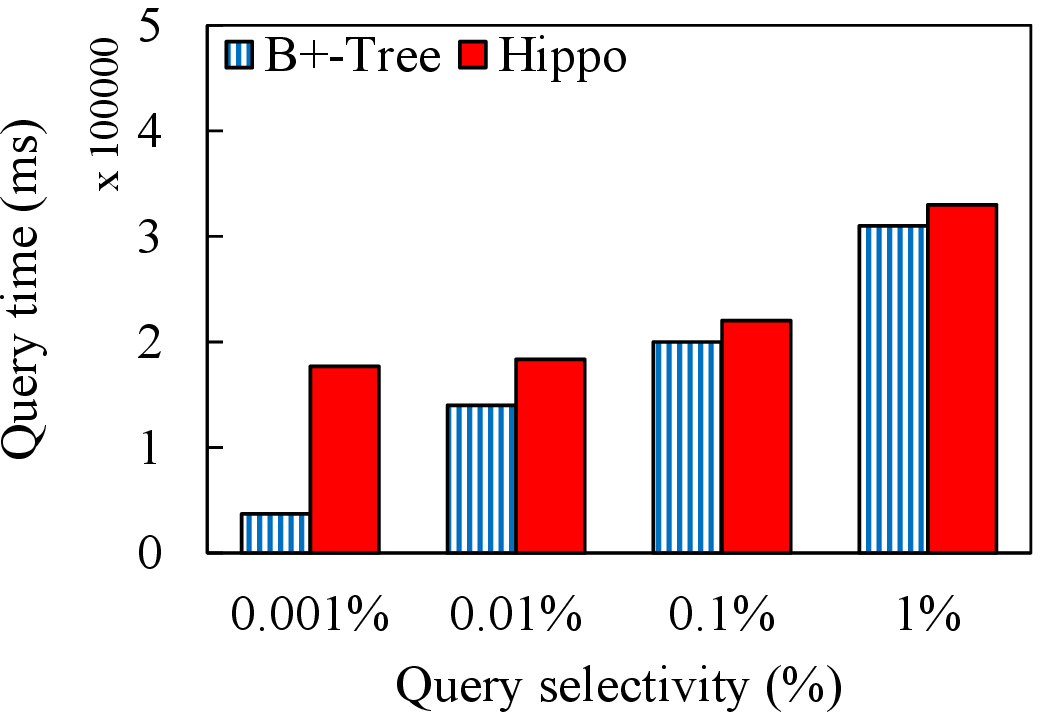}
		\caption{20 GB}
		\label{fig:mediumselectivity}
	\end{subfigure}
	~
	\begin{subfigure}{0.32\textwidth}
		\centering
		\includegraphics[width=0.95\linewidth]{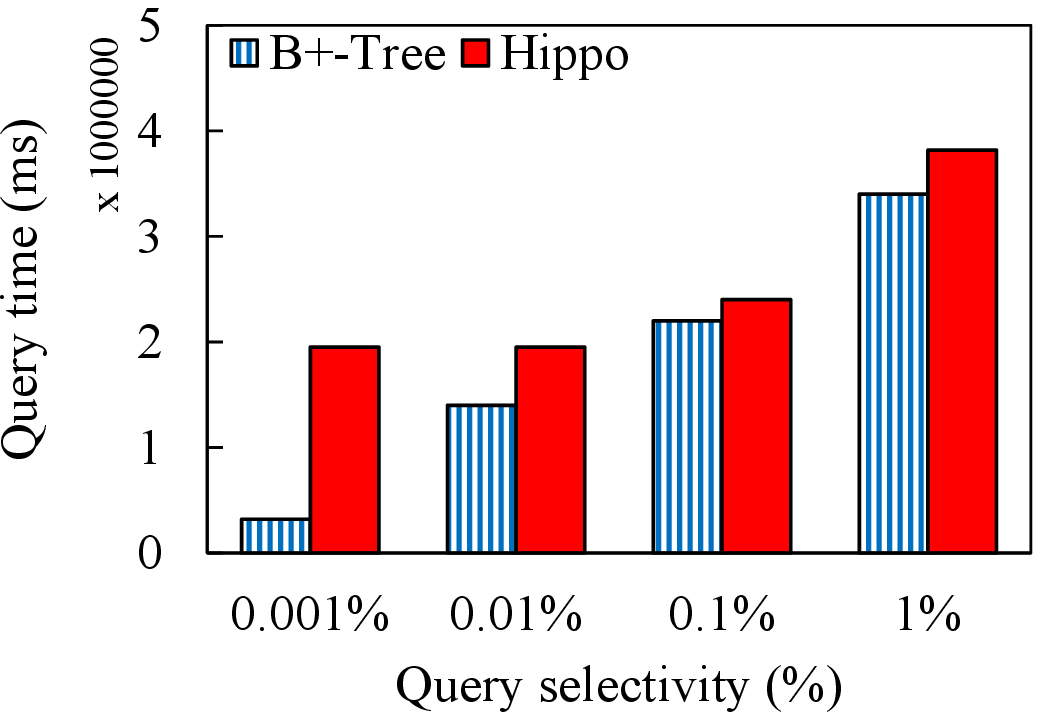}
		\caption{200 GB}
		\label{fig:bigselectivity}
	\end{subfigure}
	\caption{Index query time on different TPC-H workload size}\label{fig:selectivity}
\end{figure*}

\begin{figure*}[t]
	\centering
	\minipage{0.32\textwidth}
	\centering
	\includegraphics[width=0.95\linewidth]{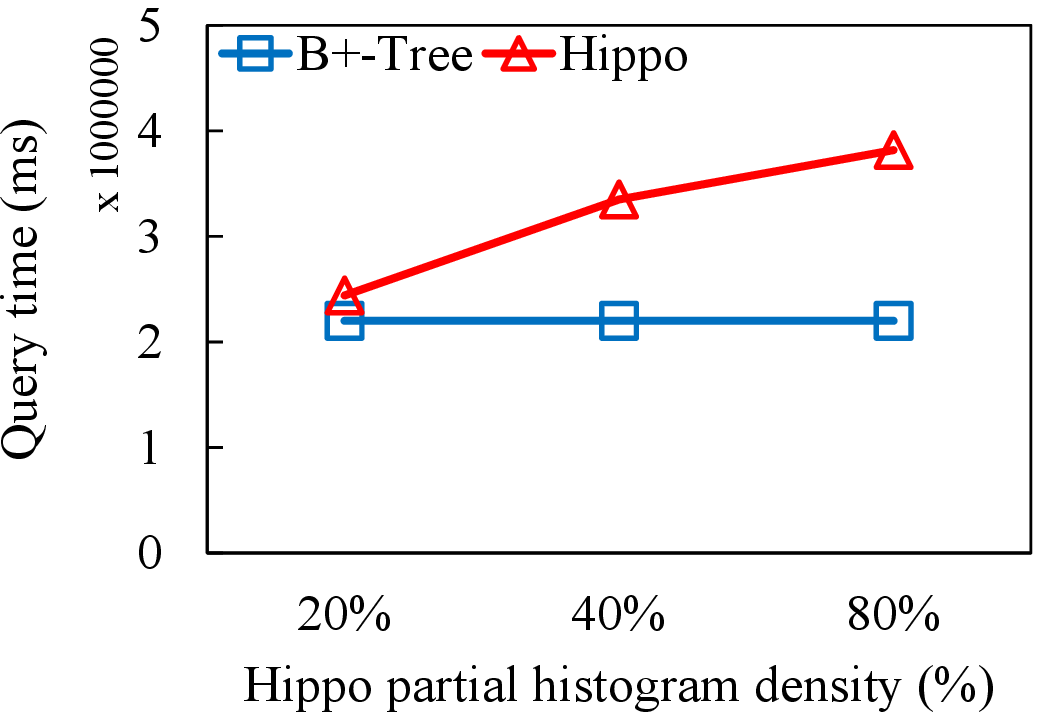}
	\caption{Partial histogram density}\label{fig:densityquery}
	\endminipage\hfill
	~
	\minipage{0.32\textwidth}
	\centering
	\includegraphics[width=0.95\linewidth]{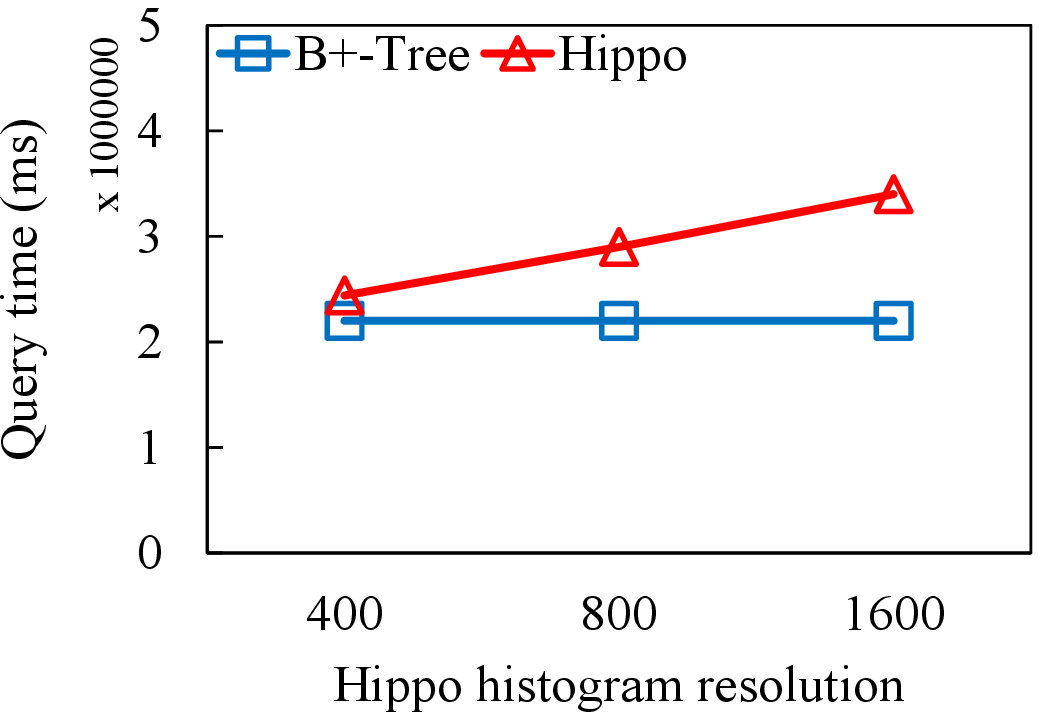}
	\caption{Histogram resolution}\label{fig:histogramquery}
	\endminipage\hfill
	~
	\minipage{0.32\textwidth}
	\centering
	\includegraphics[width=0.95\linewidth]{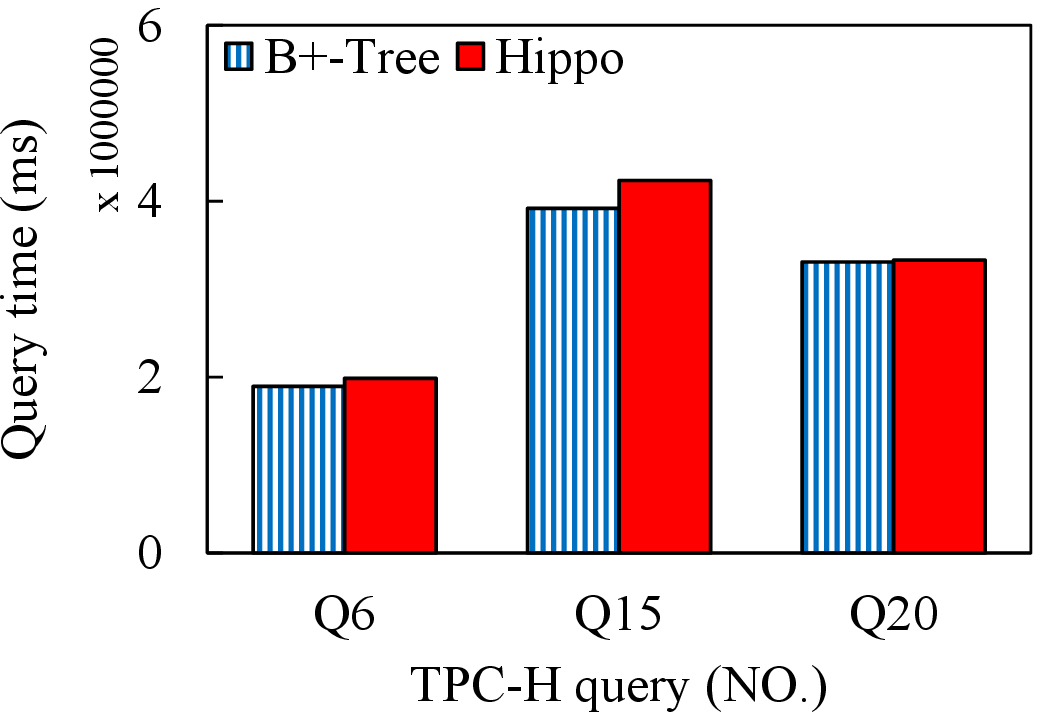}
	\caption{TPC-H standard queries}\label{fig:tpchquery}
	\endminipage
\end{figure*}

\section{Experiments}\label{experiment}

This section provides extensive experiments of Hippo along with reasonable analysis for supporting insights discussed before. For the ease of testing, Hippo has been implemented into PostgreSQL 9.5 kernel. All the experiments are completed on PostgreSQL.

{\bf Datasets and Workload.} We use TPC-H workload in the experiments with different scale factors (2, 20, 200). The corresponding dataset sizes are 2 GB, 20 GB and 200 GB. All TPC-H data follows an uniform distribution. We use the largest table of TPC-H workload - Lineitem table in most experiments and it has three corresponding sizes: 1.3 GB, 13.8 GB and 142 GB.
We compare the query time of Hippo with B$^+$-Tree through different query selectivity factors (0.001\%, 0.01\%, 0.1\% and 1\%). In addition, we also test the two indexes using TPC-H standard queries 6, 15 and 20. We use TPC-H standard refresh operation (insert 0.1\% new tuples into the DBMS) to test the maintenance overhead of B$^+$-Tree and Hippo.

{\bf Experimental Setup.} The test machine has 8 CPUs (3.5 GHz per core), 32 GB memory, and 2 TB magnetic disk with PostgreSQL 9.5 installed. Unless mentioned otherwise, Hippo sets the default partial histogram density to 20\% and the default histogram resolution to 400. The impact of parameters is also discussed.

\subsection{Implementation Details}

We have implemented a prototype of Hippo inside the core kernel of PostgreSQL 9.5 as one of the main index access methods by leveraging the underlying interfaces which include but not limited to "ambuild", "amgetbitmap", "aminsert" and "amvacuumcleanup". A database user is able to create and query a Hippo index as follows:

\begin{small}
\begin{verbatim}
CREATE INDEX hippo_idx ON lineitem USING hippo(partkey)

SELECT * FROM lineitem
         WHERE partkey > 1000 AND partkey < 2000
	
DROP INDEX hippo_idx
\end{verbatim}
\end{small}

It is also worth noting that the final Hippo implementation in PostgreSQL has some slight differences from the details above caused by some platform-dependent features as follows:

	{\bf Automatically inspect pages:} Hippo only records possible qualified page IDs in a tid bitmap format and returns it to the kernel. PostgreSQL will automatically inspect pages and check tuples against query predicates.
	
	{\bf Store the complete histogram on disk:} Compared with other disk operations, retrieving the complete histogram from PostgreSQL system cache is relatively slow so that Hippo stores it on disk and executes a binary search on it when query or update for data insertion and deletion. It is better to rebuild Hippo index if there is a huge change of the parent attribute's histogram.
	
	{\bf Vacuum tables to physically delete data:} PostgreSQL DELETE command does not really remove data from disk unless a VACUUM command is called automatically or manually. Thus Hippo will update itself for data deletion when a VACUUM command is called.

\subsection{Pre-tune Hippo parameters}

Hippo is a flexible index which can be tuned by the database user to perfectly fit his specific scenarios. There are two parameters, partial histogram density $D$ (Default value is 20\%) and complete histogram resolution $H$ (Default value is 400), discussed in this section. Referring to the estimation before, both of them have impacts on index size, initialization time, and query time. For these experiments, we build Hippo and B$^+$-tree on "partkey" attribute in Lineitem table of 200 GB TPC-H workload. As mentioned in Introduction, B$^+$-Tree has 25 GB index size at this time.

\subsubsection{Impact of partial histogram densities}
\bgroup
\def\arraystretch{1}
\begin{table}[h!]
	\small
	\centering
	\begin{tabular}{|c | c | c | c |}
		\hline
		Parameter & Value & Size & Initial. time \\ 
		\hline
		\hline
		Default  & D=20\% R=400 & 1012 MB & 2765 sec\\[1ex]
		\hline
	\multirow{2}{*}{Density (D)} &40\% & 680 MB& 2724 sec\\[1ex]
	\cline{2-4}
		&80\% & 145 MB & 2695 sec\\[1ex]
		\hline
	\multirow{2}{*}{Resolution (R)}	&800 & 822 MB & 2762 sec\\[1ex]
	\cline{2-4}
		&1600 & 710 MB & 2760 sec\\
		\hline  
	\end{tabular}
	\caption{Parameters affect Hippo indexing overhead}
	\label{table:density}
\end{table}
\egroup


Hippo introduces a terminology "partial histogram density" to dynamically control the number of pages summarized by one partial histogram. Based on the discussion before, the partial histogram density may affect Hippo size, initialization time and query time. The following experiment compares the default Hippo density (20\%) with two different densities (40\% and 80\%) and tests their query time with selectivity factor 0.1\%. According to the discussion in Section~\ref{creationestimation}, partial histograms under the three different density setting may summarize around 2 pages, 5 pages and 17 pages respectively (if one page contains 50 tuples). Thus it can be estimated that the index size of 20\% density Hippo is around 2 times of 40\% density Hippo and 8 times of 80\% density Hippo. The impact of the density on Hippo size and initialization time is described in Table~\ref{table:density} and the impact on query time is described in Figure~\ref{fig:densityquery}.

It can be observed that as we increase the density, Hippo indexing overhead decreases as expected (up to two orders of magnitude smaller than B$^+$-Tree in terms of storage) because Hippo is able to summarize more pages per partial histogram and write less index entries on disk. Similarly, Hippo which has higher density costs more query time because it is more likely to overlap with query predicates and result in more pages are selected as possible qualified pages. At this selectivity factor, Hippo which has density 20\% is just a little bit worse than B$^+$-Tree in terms of query time.

\subsubsection{Impact of histogram resolutions}


Each partial histogram of Hippo is composed of some buckets from the complete histogram. The number of buckets in this complete histogram represents the histogram resolution. The more buckets there are, the higher resolution the complete histogram has. According to the discussion before, the histogram resolution may affect index size, initialization time and query time. The following experiment compares the default Hippo histogram resolution (400) with two different histogram resolutions (800 and 1600) and tests their query time with selectivity factor 0.1\%. The density impact on the index size and initialization time is given in Table~\ref{table:density} and the impact on query time is depicted in Figure~\ref{fig:histogramquery}.

As Table~\ref{table:density} illustrates, with the growth of histogram resolution, Hippo size reduces moderately. The explanation is that Hippo which has higher histogram resolution consists of less partial histograms and each partial histogram in this Hippo may summarize more pages but the partial histogram (in bitmap format) has larger physical size because the bitmap has to store more bits.

As Figure~\ref{fig:histogramquery} shows, the query time of three Hippos varies with the growth of histogram resolution. This is because for the large histogram resolution, the query predicate may hit more buckets so that this Hippo is more likely to overlap with query predicates and result in more pages are selected as possible qualified pages. At this selectivity factor, Hippo which has histogram resolution 400 is just a little bit worse than B$^+$-Tree in terms of query time. 
 
\subsection{Compare Hippo to B$^+$-Tree}

This section compares Hippo with B$^+$-Tree in terms of indexing overhead (index size and initialization time), index maintenance overhead and index query time. To further illustrate the advantages of Hippo, we also compare these indexes using TPC-H standard queries. Hippo tested in this section uses the default setting which has histogram resolution 400 and partial histogram density 20\%.

\subsubsection{Indexing overhead}

The following experiment builds B-Tree and Hippo on attribute "partkey" in Lineitem table of TPC-H workload (2 GB, 20 GB and 200 GB) and measures their indexing overhead including index size and index initialization time. Hippo only stores disk page pointers along with their summaries so that it may have much less index entries in contrast with B$^+$-Tree. Thus it is not difficult to understand that Hippo remains an index size which is lower than B$^+$-Tree. In addition, referring to the discussion in the initialization time estimation model, Hippo initialization time should be far less than B$^+$-Tree because B$^+$-Tree has numerous nodes to be written to disk.

As Figure~\ref{fig:indexsize} illustrates, the index size increases with the growth of data size. The index size of Hippo is around 25 times smaller than that of B$^+$-Tree on all workload sizes. Thus Hippo significantly reduces the storage overhead. Moreover, as Figure~\ref{fig:indexconstructiontime} shows, Hippo index initialization is at least 1.5x faster that of B$^+$-Tree.

\subsubsection{Index maintenance overhead}

Hippo updates itself eagerly after inserting a tuple into the parent table. This eager update strategy for data insertion is also adopted by B$^+$-Tree so that the two indexes can be compared together. In terms of update time complexity, B$^+$-Tree has approximate log($Card$) and Hippo has (log(Hippo index entries) + 4). Thus it can be predicted that, for inserting same percentage of tuples, while the update time of Hippo and B$^+$-Tree is increasing with the growth of data size. Hippo will take much less time to update itself than B$^+$-Tree because $Card$ is much larger than the number of Hippo index entries. And also the difference of update time between Hippo and B$^+$-Tree will be larger on larger workload. The experiment uses TPC-H Lineitem table and creates B$^+$-Tree and Hippo on attribute "partkey". Afterwards, TPC-H refresh transaction which inserts 0.1\% new tuples into Lineitem table is executed. The insertion time of the indexes is compared in Figure~\ref{fig:insertamount}.

As Figure~\ref{fig:insertamount} shows, the two indexes take more time to update on large workload. And also the difference between B$^+$-Tree and Hippo is more obvious (1200x) on the largest workload as expected. This is because B$^+$-Tree spends much more time on searching proper tuple insert location (log($Card$)) and its update time is increasing with the growth of TPC-H workload.

Hippo updates itself lazily after deleting data which means it updates itself after many data deletions occur. In contrast, B$^+$-Tree takes an eager update strategy which has around log($Card$) update time cost. It may not make much sense to compare the two indexes for data deletion.

\subsubsection{Impact of query selectivity factors}

In this experiment, the query selectivity factors used for B$^+$-Tree and Hippo are 0.001\%, 0.01\%, 0.1\% and 1\%. According to the query time cost estimation of Hippo, the corresponding query time costs in this experiment are 0.2$Card$, 0.2$Card$, 0.2$Card$ and 0.8$Card$. Therefore, it can be predicted that there will be a great time gap between the first three Hippo queries and the last one Hippo query. On the other hand, B$^+$-Tree should be faster than Hippo at low query selectivity factor like 0.001\% but the difference between the two indexes should be narrowed with the growth of query selectivity factors.

The result in Figure~\ref{fig:selectivity} perfectly matches our predication: the last Hippo query consumes much more time than the first three queries. Among them, query time of 0.1\% selectivity factor query is a little higher than the first two because it returns more query results which costs more to retrieve. Both indexes cost more time on queries with the decreasing of query selectivity factors. B$^+$-Tree has almost similar query time with Hippo at 0.1\% query selectivity factor. It is worth noting that B$^+$-Tree consumes 25 times more storage than Hippo. Therefore, we may conclude that Hippo makes a well tradeoff between query time and index storage overhead on medium query selectivity factors like 0.1\% so that, under this scenario, Hippo is a good substitution for B$^+$-Tree if the database user is sensitive to aforementioned index overhead.

\subsubsection{TPC-H queries}

To further explore the query performance of Hippo in the real business decision support, we compare Hippo with B$^+$-Tree using TPC-H standard queries. Both of the two indexes are built on "l\_shipdate" attribute in Lineitem table of 200 GB workload. As discussed before, Hippo costs similar query time with B$^+$-Tree when the query selectivity factor is 0.1\%. Thus we find three TPC-H queries which have typical range queries on "l\_shipdate" attribute (Query 6, 15 and 20) and set the range query selectivity factor to 0.1\% which means one week. The query plans of the three queries are described as follows:

{\bf Query 6} This query has a very simple plan. It firstly performs an index search on Lineitem table using one of the candidate indexes, then filters the returned values and finally aggregates the values to calculate the result.

{\bf Query 15} This query builds a sub-view beforehand and embeds it into the main query twice. The range query which leverages the candidate indexes is a part of the sub-view.

{\bf Query 20} The candidate indexes are invoked in a sub-query. Then range query results are sorted and aggregated for calculation. The result is cached into memory and used the upper level query.

As Figure~\ref{fig:tpchquery} depicts, Hippo consumes similar query time with B$^+$-Tree on Query 6, 15 and 20. The difference between the two indexes is more obvious on Query 15 because this query invokes the range query twice. Therefore, we may conclude that Hippo may achieve almost similar query performance with B$^+$-Tree at the 25 times smaller storage overhead when the query selectivity factor is 0.1\%.

\def\arraystretch{1}
\begin{table}[t]
	\scriptsize
	\centering
	\begin{tabular}{|c | c | c | c | c|}
		\hline		
		{Index} & {Fast} & Guaranteed  & Low & Fast \\ 
		{type} & {Query} & Accuracy & {Storage} & {Maintenance} \\
		\hline \hline
		{B$^+$-Tree} & {\cmark} & {\cmark} & {\xmark} & {\xmark}\\[1ex]
		\hline		
		{Compressed} & {\xmark} & {\cmark} & {\cmark} & {\xmark}\\[1ex]
		\hline		
		{Approximate} & {\cmark} & {\xmark} & {\cmark} & {\xmark}\\[1ex]
		\hline		
		{Sparse} & {\xmark} & {\cmark} & {\cmark} & {\cmark}\\[1ex]
		\hline		
		Hippo & {\cmark} & {\cmark} & {\cmark} & {\cmark}\\
		\hline		
	\end{tabular}
	\caption{Compared Indexing Approaches}
	\label{table:relatedwork}
\end{table}

\section{Related work}
\label{relatedwork}

Table~\ref{table:relatedwork} summarizes state-of-the-art database index structures in terms of query time, accuracy, storage overhead and maintenance overhead.

{\bf Tree Index Structures:} B$^+$-Tree is the most commonly used type of indexes. The basic idea can be summarized as follows: For a non-leaf node, the value of its left child node must be smaller than that of its right child node. Each leaf node points to the physical address of the original tuple. With the help of this structure, searching B$^+$-Tree can be completed in one binary search time scope. The excellent query performance of B$^+$-Tree and other tree like indexes is benefited by their well designed structures which consist of many non-leaf nodes for quick searching and leaf nodes for fast accessing parent tuples. This feature incurs two inevitable drawbacks: (1) Storing plenty of nodes costs a huge chunk of disk storage. As shown in Section~\ref{introduction}, it results in non-ignorable dollar cost and huge initialization time in big data scenarios. (2) Index maintenance is extremely time-consuming. For any insertions or deletions occur on parent table, tree like indexes firstly have to traverse themselves for finding proper update locations and then split, merge or re-order one or more nodes which are out of date.



{\bf Compressed Index Structures:} Compressed indexes try to drop some repeated index information as much as possible beforehand for saving space and recover it as fast as possible upon queries from users but they all have guaranteed query accuracy. These techniques are applied to tree indexes~\cite{GRS98,GCC+14} and bitmap indexes~\cite{FSV10,LKA10,SW06,ZHN+06} (low cardinality and read-only datasets). Though compressed indexes are storage economy, they require additional time for compressing beforehand and decompressing on-the-fly. Compromising on the time of initialization, query and maintenance is not desirable in many time-sensitive scenarios. Hippo on the other hand reduces the storage overhead by dropping redundancy tuple pointers and hence still achieves competitive query response time. 


{\bf Approximate Index Structures:} Approximate indexes~\cite{AA14,HS05,SYU+00} give up the query accuracy and only store some representative information of parent tables for saving indexing and maintenance overhead and improving query performance. They propose many efficient statistics algorithms to figure out the most representative information which is worth to be stored. In addition, some people focus on approximate query processing (AQP)\cite{AMK+14,ZGM+14} which relies on data sampling and error bar estimating to accelerate query speed directly.  However, trading query accuracy makes them applicable to limited scenarios. On the other hand, Hippo, though still reduces the storage overhead, only returns exact answer that match the query predicate.

{\bf Sparse Index Structures:} Sparse index (denoted as Zone Map Index in IBM Data Warehouse\cite{BZ98}, Data Pack structure in Infobright\cite{SE09}, Block Range Index in PostgreSQL\cite{SR86}, and Storage Index in Oracle Exadata\cite{W12}) is a simple index structure implemented by many popular DBMS in recent years. Sparse index only stores pointers which point to disk pages of parent tables and value ranges (min and max values) in each page so that it can save indexing and maintenance overhead. It is generally built on ordered attributes. For a posed query, it finds value ranges which cover or overlap the query predicate and then rapidly inspects the associated few parent table pages one by one for retrieving truly qualified tuples. However, for most real life attributes which have unordered data, sparse index has to spend lots of time on page scanning because the stored value ranges (min and max values) may cover most query predicates and encumber the page inspection. Therefore, an efficient yet concise page summarizing method (i.e., Hippo) instead of simple value ranges is highly desirable.


\section{Conclusion}
\label{conclusion}

The paper introduces Hippo a sparse indexing approach that efficiently and accurately answers database queries while occupying up to two orders of magnitude less storage overhead than de-facto database indexes, i.e., B$^+$-tree. To achieve that, Hippo stores pointers of pages instead of tuples in the indexed table to reduce the storage space occupied by the index. Furthermore, Hippo maintains histograms, which represent the data distribution for one or more pages, as the summaries for these pages. This structure significantly shrinks index storage footprint without compromising much on performance of common analytics queries, i.e., TPC-H workload. Moreover, Hippo achieves about three orders of magnitudes less maintenance overhead compared to the B$^+$-tree. Such performance benefits make Hippo a very promising alternative to index data in big data application scenarios. Furthermore, the simplicity of the proposed structure makes it practical for database systems vendors to adopt Hippo as an alternative indexing technique. In the future, we plan to adapt Hippo to support more complex data types, e.g., spatial data, unstructured data.
\newpage
\bibliographystyle{abbrv}
\bibliography{hippo}  
\end{document}